\documentclass[conference]{IEEEtran}
\usepackage{graphicx} 
\usepackage{cite}
\usepackage{amsmath,amssymb,amsfonts}
\usepackage{algorithmic}
\usepackage{graphicx}
\usepackage{textcomp}
\usepackage[normalem]{ulem}
\usepackage{xcolor}
\usepackage{booktabs}
\usepackage{tikzsymbols}
\usepackage{url}
\usepackage{tikz}
\usepackage{multirow}
\usepackage{pifont}
\usepackage{pgfplots} 
\pgfplotsset{compat=1.18} 
\usepackage{xfrac} 
\usepackage{adjustbox}
\usepackage{makecell}

\usepackage[draft,authormarkup=none]{changes} 

\definechangesauthor[name={Hanieh TotonchiAsl}, color=blue]{HTA}
\definechangesauthor[name={Danella}, color=red]{DZh}

\begin{document}

\IEEEoverridecommandlockouts

\IEEEpubid{\makebox[\columnwidth]{Accepted to special session track.\hfill} \hspace{\columnsep}\makebox[\columnwidth]{ }}

\title{Interflow: Breaking the Offline–Online Barrier for Private AI at the Edge}
\author{
    \IEEEauthorblockN{Hanieh Totonchi Asl, 
    Ebrahim Nouri, 
    Zhuoran Li, 
    Danella Zhao}
    \IEEEauthorblockA{Department of Electrical and Computer Engineering, University of Arizona, 
    Tucson, United States\\
    \{haniehta, ebinouri, zli1122, danellazhao\}@arizona.edu\\}
}

\maketitle

\begin{abstract}
Privacy-preserving machine learning (PPML) offers a promising path toward bringing powerful AI services to the edge without exposing sensitive user data.
However, its practical deployment remains constrained by the high communication, computation, and data-movement costs of secure multi-party computation (MPC), particularly during nonlinear inference. 
Existing MPC systems typically decouple input-independent correlated randomness generation from online execution, creating substantial memory pressure to buffer single-use cryptographic data. Moreover, limited storage can force a stop-and-refill execution pattern that repeatedly stalls the pipeline to replenish these buffers, bringing randomness generation back onto the critical path.
We propose 
\textit{Interflow}, an algorithm--hardware
co-design that rethinks this separation by 
batch-wise processing and pipelining correlated-randomness generation with online execution: while the current batch consumes pre-generated randomness to execute secure nonlinear operations, 
correlation generation proceeds concurrently for the next batch. By leveraging interleaved execution, 
efficient data movement, and hardware acceleration,
the design transforms preprocessing from a serialized, storage-intensive offline phase into a continuously overlapped pipeline that generates correlated randomness closer to the point of consumption, reducing memory footprint and data movement overhead.
Implemented on the Zynq UltraScale+ ZCU102, \textit{Interflow} improves nonlinear evaluation throughput by up to $1.55\times$ for CNN operators and $1.48\times$ for transformer operators.
It achieves up to $1.26\times$ speedup for ResNet-50 and up to $1.21\times$ speedup for BERT-base, and reduces memory footprint by up to $37\%$,
compared with 
baseline 
frameworks.
\end{abstract}

\IEEEpeerreviewmaketitle

\section{Introduction}


Machine Learning as a Service (MLaaS) enables users to outsource inference to the cloud; however, its deployment in sensitive domains such as healthcare, finance, and personal services raises fundamental privacy concerns. Privacy-preserving machine learning (PPML) addresses this challenge through cryptographic primitives such as homomorphic encryption (HE) and secure multi-party computation (MPC), enabling computation over encrypted data without exposing raw inputs and model weights~\cite{juvekar2018gazelle, riazi2019xonn, rouhani2018deepsecure, mishra2020delphi, hecnn2023}. 
Practical two-party PPML frameworks commonly adopt a hybrid approach, evaluating 
linear layers via HE and 
nonlinear operations (such as ReLU, GELU, MaxPooling, and Softmax) through secret-sharing (SS)-based MPC protocols~\cite{rathee2020cryptflow2, huang2022cheetah, pang2024bolt, lu2025bumblebee, zhang2024individual}. 
Despite their benefits, 
secure nonlinear operations incur substantial computation and communication overhead and often require large amounts of correlated randomness, making them a critical bottleneck on resource-constrained edge devices~\cite{li2025silentflow, tamimpc2026}.

Modern MPC protocols adopt an offline–online paradigm, in which input-independent cryptographic correlations such as oblivious transfers (OTs) and Beaver triples are generated during an offline preprocessing phase, shifting much of the expensive computation and communication away from the 
online phase~\cite{huang2022cheetah, rathee2020cryptflow2}. 
Pre-generating correlated randomness requires substantial storage; in practice, systems maintain only a bounded pool of precomputed OTs or Beaver triples (e.g., $10^{7}$ OTs in Cheetah). 
Once exhausted, regeneration stalls execution and introduces significant latency, bringing offline preprocessing onto the online critical path. This bottleneck becomes especially severe on resource-constrained devices.

\textbf{Key Observations.} 
Conventional preprocessing treats correlation generation as a decoupled phase. Under resource constraints, however, correlated randomness becomes a continuous lifecycle challenge: fresh values must be generated, stored in different representations and widths, converted for protocol-specific operations, delivered to their consumption points, and discarded after a single use. 
Our characterization of a baseline framework on an edge client shows that repeated correlation regeneration accounts for 43\% of client-side secure-comparison cycles during ResNet-50 inference. This substantial overhead motivates optimizing the entire correlation lifecycle, as accelerating generation alone may simply shift the bottleneck to storage, conversion, or data movement rather than eliminating it.

\textbf{Design Insights.}
Based on these observations, we present \emph{Interflow}, an algorithm--hardware co-design for OT-based secure nonlinear inference that jointly manages the complete correlation 
lifecycle, from generation and representation-aware storage to sender/receiver-specific conversion and online consumption.
Our central insight is to 
split
preprocessing at its final communication dependency and consumption boundary. 
This allows receiver-local COT generation to be deferred until near consumption and integrated directly into the execution dataflow. Building on this decomposition, Interflow employs batch-wise, role-aware processing to overlap correlation generation, conversion, and consumption, avoiding the need to generate and store the full set of 
expanded correlations in advance.


Our main contributions are summarized as follows:
\begin{itemize}
    \item \textbf{System-level characterization.} We identify and quantify correlation-lifecycle bottlenecks in OT-based nonlinear inference on resource-constrained edge devices, including correlation regeneration on the online critical path, wide intermediate representations, and protocol-imposed sender--receiver asymmetry.
    \item 
    \textbf{Deferred correlation processing with cross-phase dataflow.} 
    We propose a role-aware COT extension design that defers receiver-local computation beyond communication-dependent processing and integrates correlation generation and conversion with online secure execution in a memory-bounded, batch-streaming dataflow, reducing critical-path regeneration, memory footprint, and data movement.
    \item \textbf{Role-aware preprocessing.} 
    We design a dual-path correlation-processing architecture that coordinates bidirectional OT for Beaver triple generation 
    to improve throughput, reduce intermediate buffering, and enable efficient streaming execution. 
    We further reformulate sender-side correlation
    conversion as independent index-aligned hash streams to enable parallel execution.
    \item 
    \textbf{Practical edge PPML evaluation.}
    We implement Interflow on a Zynq UltraScale+ ZCU102 platform and 
    evaluate secure nonlinear inference for ResNet-50 and BERT-base.
    Interflow improves nonlinear evaluation throughput by up to~$1.55\times$ for CNN operators and~$1.48\times$ for transformer operators, achieves up to $1.26\times$ speedup for ResNet-50 and $1.21\times$ speedup for BERT-base, and reduces memory footprint by up to $37\%$,
    demonstrating the practicality of efficient privacy-preserving AI inference on resource-constrained edge platforms.
\end{itemize}


\section{Preliminaries and Background}\label{Sec:background}
\subsection{Fundamental MPC primitives for Nonlinear Evaluation}\label{Sec:back_nonlin}

\begin{figure}[t] 
    \centering 
    \includegraphics[width=\linewidth, trim={0cm 0.55cm 0cm 0cm}, clip ]{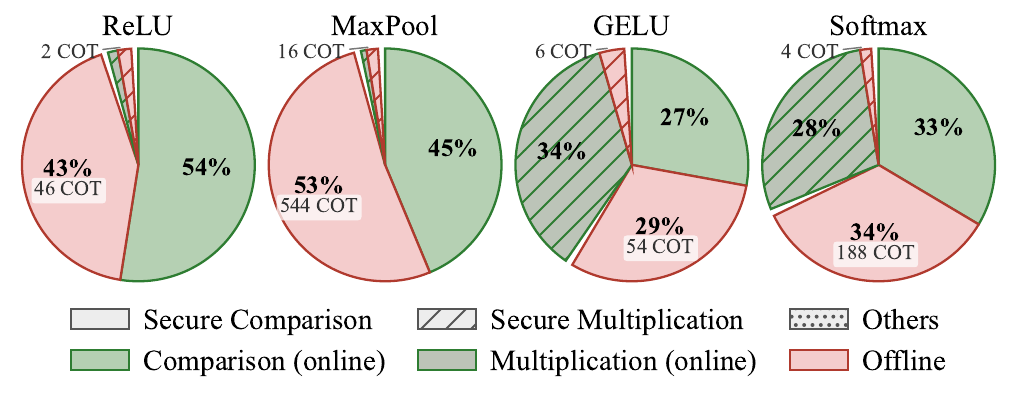} 
    \vspace{-15pt}
    \caption{Breakdown of representative secure nonlinear operators into comparison, multiplication, and other primitives. 
    Correlation volumes are reported per output element. Operators are measured in isolation, x86$\leftrightarrow$x86 over a 1 Gbps/0.3 ms link, at $n_t{=}4$ threads: ReLU and MaxPool under Cheetah at ResNet-50 dimensions, GELU and Softmax under BumbleBee at BERT-base dimensions. BumbleBee's multiplications draw a few correlations because they are evaluated homomorphically (RLWE) rather than with arithmetic Beaver triples.} 
    \label{fig:nonlinear_decomposition} 
    \vspace{-15pt}
\end{figure}

MPC-based nonlinear inference can be largely decomposed into two fundamental secure primitives: comparison and multiplication. Fig.~\ref{fig:nonlinear_decomposition} summarizes how representative nonlinear operations in 
Cheetah~\cite{huang2022cheetah} and BumbleBee~\cite{lu2025bumblebee} are decomposed into secure computation primitives. 
ReLU and MaxPooling rely primarily on secure comparisons (${\sim}97\%$),
followed by secure multiplication or selection to produce the corresponding outputs. 
More complex nonlinearities, such as GELU and Softmax, are typically evaluated via piecewise or polynomial approximations 
that similarly decompose their execution into secure comparisons (${\sim}56{-}67\%$) and multiplications (${\sim}29{-}37\%$).
These two primitives therefore constitute key building blocks of secure nonlinear inference: comparison 
resolves signs, ordering, and selection predicates, while multiplication governs
arithmetic evaluation and masked selection.
Their frequent execution makes
the associated correlation generation and interactive execution major sources of computation, communication, and storage overhead.

Despite their different functionality, these primitives rely heavily on a common class of input-independent correlated randomness
to lighten online execution: 
OTs and Boolean triples for secure comparison and secure AND operations, and arithmetic triples for secure multiplication. 

\textbf{Secure comparison.}\label{Sec:back_comp}
To evaluate $\mathsf{1}(x>y)$, the protocol employs the OT-based Millionaires' protocol with radix decomposition~\cite{rathee2020cryptflow2, huang2022cheetah}. 
For radix $2^r$, each $w$-bit operand is partitioned into $d = \lceil w / r \rceil$ digits.
During online execution, the sender evaluates the comparison outcomes ($\mathsf{1}(x^{(i)}>M)$) for every $M\in[0,2^r)$ as the messages of a $1$-out-of-$2^r$ OT, while the receiver selects the message indexed by its private digit $y^{(i)}$.
The resulting digit-level comparison and equality shares are subsequently aggregated from the most significant digit to the least significant digit, 
producing an XOR share of the comparison result. 

Digit aggregation requires secure AND operations over XOR-shared intermediate bits. Given $u{=}u_0\oplus u_1$ and $v{=}v_0\oplus v_1$, preprocessing supplies an XOR-shared Boolean Beaver triple $\langle a\rangle,\langle b\rangle,\langle c\rangle$ satisfying $c=a\land b$. During online execution, each party locally computes $e_j=u_j\oplus a_j$ and $f_j=v_j\oplus b_j$; the parties exchange these shares to reconstruct the masked values $e$ and $f$, and then combine them with their local triple shares to obtain an XOR share of $u\land v$. Thus, each secure AND consumes one fresh Boolean triple and requires one masked value exchange followed by local Boolean operations.


\textbf{Secure multiplication.}
Secure multiplication applies a similar Beaver-masking principle as secure AND, over the ring $\mathbb{Z}_{2^w}$.
Preprocessing supplies additive shares of one fresh arithmetic Beaver triple $(a,b,c)$ satisfying $c{=}a\cdot b$. Given additive shares of $x$ and $y$, the parties form and exchange shares of $e{=}x-a$ and $f{=}y-b$, and locally derive shares of $xy=c+af+be+ef$. 

Although correlation generation is independent of the private operands, online comparison cannot proceed until the corresponding correlations are available.
Across complete nonlinear operators, repeated comparison and multiplication thus create a producer--consumer boundary between correlation preprocessing and input-dependent execution, whose storage and refill costs are quantified in Section~\ref{Sec:motivation}.

\begin{figure*}[t]
    \centering
    \includegraphics[width=\linewidth, trim={0.0cm 0.4cm 0.2cm 0.2cm}, clip]{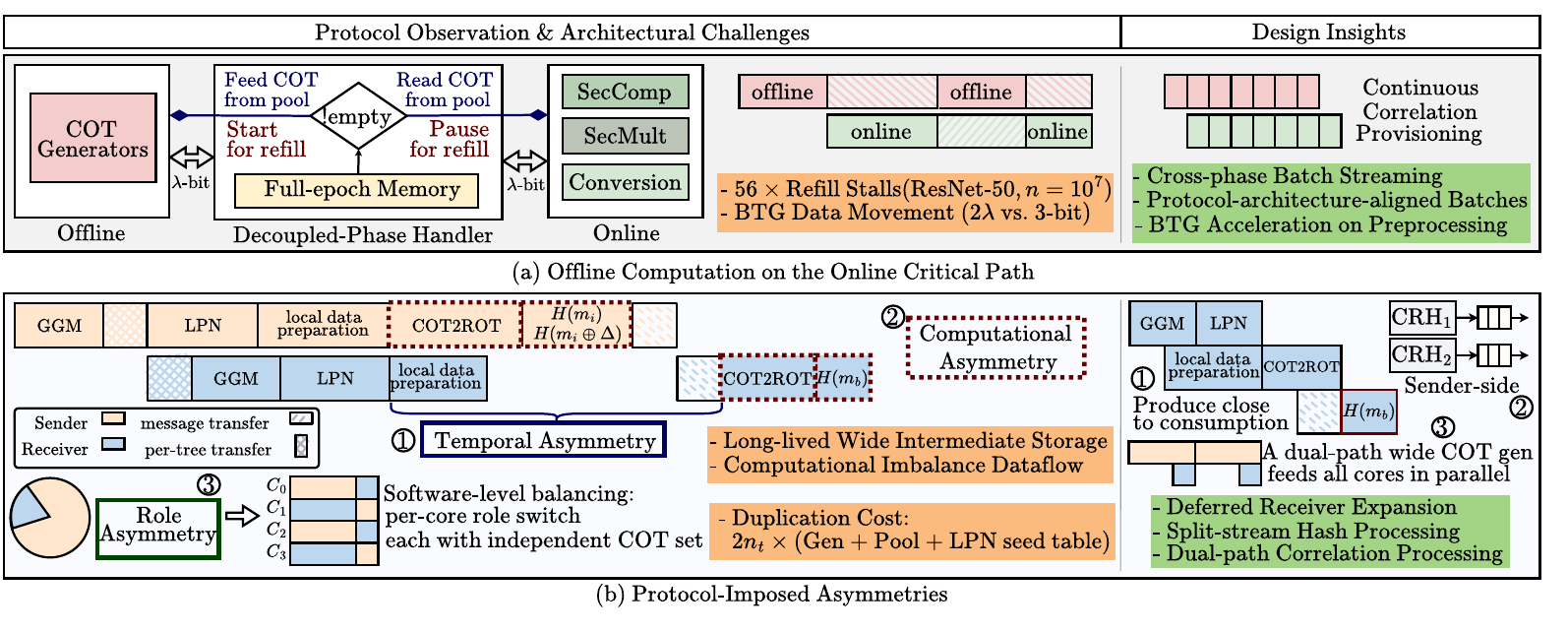}
    \vspace{-20pt}
    \caption{From protocol observations to architectural design principles.}
    \vspace{-15pt}
    \label{fig:motivation}
\end{figure*}

\subsection{Correlation Processing}\label{Sec:back_corr}
The nonlinear primitives above consume three closely related types of correlated randomness as their masking material. 
In our setting, correlated OTs (COTs) serve as the underlying correlation source that is converted into protocol-specific forms, including random OTs (ROTs) and Beaver triples, before one-time 
online consumption. 
ROTs provide randomized OT instances used by comparison-oriented protocols, and Beaver triples provide multiplication-ready correlations.

\textbf{Correlated Oblivious Transfer.}\label{Sec:back_COT}
In one COT instance, the sender holds a 
uniformly distributed $\lambda$-bit value $COT_s$ and a global correlation $\Delta$, while the receiver holds a 
choice bit $b_i$ and 
$COT_r{=}COT_s\oplus b_i\Delta$.
Ferret's silent correlation generator produces these COTs in two stages~\cite{boyle2019silentot,yang2020ferret}.
First, a GGM-based protocol expands each tree of height $h$ into $2^h$ leaves using compact level-wise communication and local tree expansion. Second, an element-wise LPN computation combines the sparse tree outputs with random publicly indexed entries from a seed vector to produce the final COTs, while retaining a subset for the next extension epoch.

\textbf{Random Oblivious Transfer.}\label{Sec:back_rot}
Raw COT values retain the linear relation induced by $\Delta$ and do not directly serve as independent message pads. 
A correlation-robust hash (CRH) converts them into ROT outputs~\cite{huang2022cheetah, Guo2020mitccrh}, using the Matyas--Meyer--Oseas (MMO) feed-forward construction,   
$H(x) {=} \pi(x) \oplus x$,
where $\pi$ is commonly instantiated as fixed-key AES for $\lambda=128$ security setting.

\textbf{Beaver Triple.}\label{Sec:back_BTG}
A Boolean Beaver triple consists of XOR shares $(a_j,b_j,c_j)$ satisfying $c_0\oplus c_1=(a_0\oplus a_1)\land(b_0\oplus b_1)$.
Each party computes its same-party product locally, while two opposite-direction ROTs provide shares of the cross-products.
In the first direction, $P_0$ acts as the ROT sender with pads $(ROT_s^0,ROT_s^0\oplus a_0)$, while $P_1$ acts as the receiver and selects with $b_1$ to obtain $ROT_r^1{=}ROT_s^0\oplus(a_0\land b_1)$.
The second direction reverses the roles to share $a_1\land b_0$.
Each party then locally computes its final outputs as:
$c_j = (a_j\land b_j)\oplus ROT_s^j\oplus ROT_r^j$,
yielding $c_0\oplus c_1=a\land b$. 
Thus, each party acts once as ROT sender and once as ROT receiver. 
Generating one triple consumes two ROTs, one per direction, and requires three CRH evaluations at each party: two for its sender-role pad pair and one for its receiver-role selected pad.
\section{Motivation and Challenges}\label{Sec:motivation}

While secure protocols reduce online execution overhead through input-independent preprocessing, their implementation on resource-constrained platforms exposes important protocol–architecture mismatches.
Fig.~\ref{fig:motivation} highlights two bottlenecks we observe in secure nonlinear inference.
First, finite correlation storage couples preprocessing back to online execution, causing regeneration to stall the execution pipeline when precomputed correlations are exhausted. 
Second, the asymmetric sender and receiver roles exhibit different computation and data-lifetime requirements, leading to inefficient resource utilization and correlation processing when handled uniformly. 
We quantify their impact and identify corresponding design opportunities through profiling of Cheetah~\cite{huang2022cheetah} on ResNet-50 inference with 43-bit secure comparisons.


\subsection{Offline Computation on the Online Critical Path}\label{sec:motivation_refill} 

Modern nonlinear protocols adopt an offline–online paradigm that moves expensive, input-independent correlation generation off the input-dependent execution path, reducing online computation and communication. 
This approach typically pre-generates correlations in large batches and buffers them for subsequent online consumption, trading online cost for storage, buffering, and data movement (Fig.~\ref{fig:motivation} (a)). 
Our profiling of Cheetah, however, reveals a fundamental tension between correlation storage and regeneration. 
Because each correlation is consumed only once, the buffer must be continuously replenished as inference proceeds. 

On resource-constrained edge devices, limited memory forces smaller batches, turning correlation processing into a continuous lifecycle rather than a one-time offline task: fresh COTs must be generated batch-by-batch, stored in intermediate representations, converted into operation-specific forms, delivered to their consumption points, and discarded after one-time online use. 
Therefore, correlation exhaustion 
induces a
recurring \textbf{stop-and-refill} pattern that brings preprocessing directly onto the online critical path, accounting for $43\%$ of the client's secure-comparison cycles (Table~\ref{tab:refil_stall}).

This observation exposes a key design opportunity: rather than treating correlation generation, conversion, and consumption as separate phases, their lifecycle can be reorganized as a batch-streaming dataflow that
overlaps correlation preparation with online secure execution, thereby reducing storage and data movement while hiding correlation-processing latency.


\begin{table}[t]
    \centering
    \vspace{-15pt}  
    \caption{COT regeneration overhead 
    for Cheetah ResNet-50 inference.}
    \vspace{-5pt}    
    \setlength{\tabcolsep}{3pt}
    \resizebox{0.49\textwidth}{!}{%
    \begin{tabular}{lrrrrrr}
    \toprule
    Platform & Pool Size & Refills & Stall/Refill & \shortstack{Online Stall} & \shortstack{Cmp Online Stall} & Latency \\
    \midrule
    x86     & 1,439.7 MB & 56    & 0.33 s  & 4.0 \%  & 17.5 \% & 379.6 s \\
    ZCU102   & 118.9 MB   & 2,269 & 0.165 s & 21.7 \% & 43.0 \% & 455.0 s \\
    \bottomrule
    \end{tabular}
    }
    \vspace{-10pt}
    \label{tab:refil_stall}
\end{table}

\subsection{Protocol-Imposed Asymmetries} \label{sec:motivation_role_asymmetry} 
Secure nonlinear inference processes correlated randomness through role-dependent protocol steps. 
Although correlations can be pre-generated, their online production and consumption expose three architectural asymmetries: temporal, computational, and role assignment (Fig.~\ref{fig:motivation} (b)).


\textbf{Temporal Asymmetry.}  
Sender- and receiver-side correlations are consumed at different points in the protocol execution path.
As shown in Fig.~\ref{fig:motivation}\textcircled{1}, 
sender-side masking and transmission must complete before receiver-side payload recovery can occur, even under concurrent execution of different OT instances and preprocessing batches. 
A sender pad remains live until the application message is masked, whereas a receiver pad remains live until the corresponding masked payload arrives for recovery. 
Pre-generating full-epoch correlations inflates the memory footprint across execution phases. While streaming generation can reduce the buffer size, it cannot alter these protocol dependencies; the memory subsystem must therefore incorporate decoupled, role-aware, index-aligned buffers to manage asymmetric correlation lifespans.

\textbf{Computational Asymmetry.}  
As shown in Fig.~\ref{fig:motivation}\textcircled{2}, COT-to-ROT conversion also assigns imbalanced local computation to the two roles.
For a given OT index $i$, the sender evaluates the CRH on both correlated branches ($H_i(q_i)$ and $H_i(q_i\oplus\Delta)$), whereas the receiver evaluates only the selected branch $H_i(q_i\oplus b_i\Delta)$. 
This compute asymmetry manifests after raw valid COTs become available, 
constraining performance regardless of how the correlation generator is structured, whether implemented as
a prefilled memory pool, a background refill engine, or a streaming generator.

\textbf{Role Asymmetry.} 
Composing role-dependent primitives into a nonlinear operator creates an application-level dataflow asymmetry between the two parties.
In the single-core Cheetah ResNet-50 profile, the client consumes $416.0$\,M receiver-role COTs and $99.4$\,M sender-role COTs, establishing a $4.18\times$ imbalance. 
As shown in Fig.~\ref{fig:motivation}\textcircled{3}, the baseline
software implementation covers this imbalance by mirroring party roles across alternating worker threads, but at the cost of replicating complete generator instances.
With $n_t$ online consumers, each party maintains $2n_t$ independent COT generators, each with a dedicated COT buffer and LPN seed table. 
At $n_t{=}4$, these instances occupy $1,445.6$\,MB, or $56.3\%$ of the client's measured peak memory footprint.
Concurrent accesses to the replicated seed tables further exacerbate the cache pressure identified in Section~\ref{sec:motivation_refill}.
This state duplication exposes a hardware opportunity for memory optimization.

Together, the three asymmetries 
highlight potential headroom for improvement beyond scaling the cryptographic kernel alone.
The architecture must coordinate when correlations are generated, how role-specific data are buffered, and how sender/receiver processing capacity is activated. 
By exploiting these opportunities, 
Interflow coordinates the downstream lifecycle of each correlation from producer output through role-specific conversion to online consumption.




\begin{figure}[t]
    \centering
    \makebox[\columnwidth]{%
        \includegraphics[width=0.75\linewidth, trim={0.7cm 0.5cm 0.7cm 0.75cm}, clip]{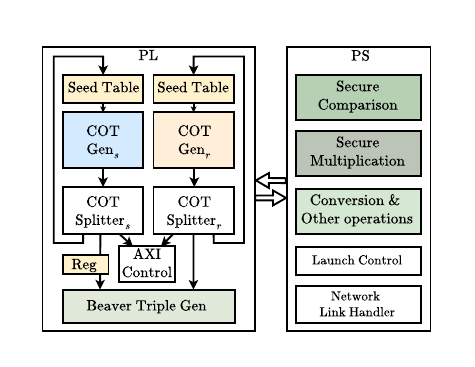}
    }
    \vspace{-20pt}
    \caption{System architecture of \emph{Interflow} illustrating the PS--PL partition.}
    \vspace{-15pt}
    \label{fig:system}
\end{figure}
\section{Proposed Design}\label{Sec:design}
Throughout, we consider the standard semi-honest (honest-but-curious) two-party setting used by CryptFlow2~\cite{rathee2020cryptflow2} and Cheetah~\cite{huang2022cheetah}, in which both parties follow the specified protocol, 
and neither 
misuses correlated primitives.
\vspace{-5pt}

\subsection{Design Overview}\label{Sec:design_overview}
Fig.~\ref{fig:system} presents the proposed hardware--software architecture for secure nonlinear inference. The design partitions execution between the processing system (PS) and the programmable logic (PL) according to the protocol's computational and communication characteristics. 
The PL accelerates the compute-intensive input-independent preprocessing kernels, including COT generation, COT-to-ROT conversion, and Beaver triple generation. 
At the same time, the PS executes the interactive input-dependent online phase, including protocol coordination, communication with the remote party, and lightweight local operations. The PS also configures and launches the accelerator and manages the AXI interface through which it exchanges intermediate streaming data with the PL. Correlations are tagged by logical OT direction and batch so that subsequent stages preserve their protocol association and consumption order. This organization integrates hardware-accelerated preprocessing computation, inter-party communication, and shared-memory data movement within a unified execution flow.

To improve pipeline utilization and reduce inter-stage stalls, the PL employs a stage-balanced streaming architecture in which each stage begins processing as soon as its input becomes available. Separate correlation streams are maintained for the sender and receiver roles, with correlation generation scheduled according to the demands of upcoming nonlinear operations. 
The design further overlaps online processing of the current batch with generating and converting correlated randomness for the next batch. Consequently, successive pipeline stages operate concurrently, and the steady-state processing rate is governed primarily by the slowest stage rather than by the sum of their execution times. This producer-consumer organization reduces the volume of pending intermediate state and correlation-generation and refill delays from the secure inference pipeline.

\subsection{Role-aware COT Extension Deferral}\label{Sec:design_deferral}
\begin{figure}[t]
    \centering
    \makebox[\columnwidth]{%
        \includegraphics[width=0.9\linewidth, trim={0.7cm 0.8cm 0.8cm 0.55cm}, clip]{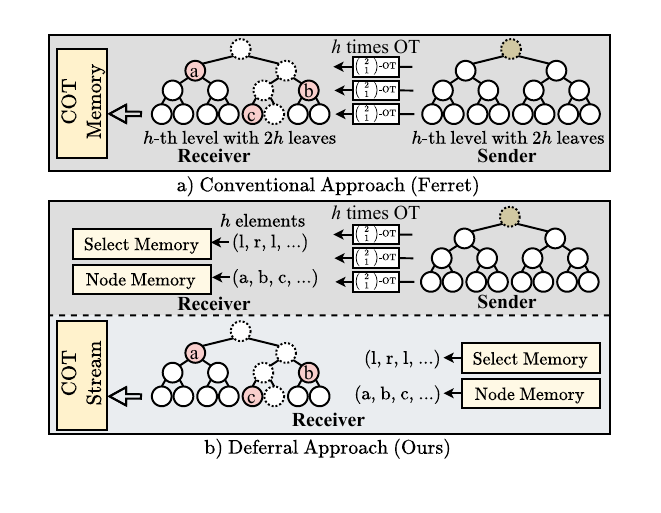}
    }
    \vspace{-20pt}
    \caption{Decoupling communication and computation on the receiver side to defer its COT expansion closer to its consumption step, storing only $h$ selected nodes and one-bit child choice per tree instead of all $2^h$ leaves.}
    \vspace{-15pt}
    \label{fig:Protocol}
\end{figure}

Motivated by the temporal asymmetry identified in Section~\ref{sec:motivation_role_asymmetry}, we propose \emph{Role-Aware Deferred COT Extension}, 
which separates interactive preprocessing from receiver-side local computation. 
As shown in Fig.~\ref{fig:Protocol}, the sender follows the conventional expansion flow required to construct the protocol messages, whereas the receiver keeps a compact record and defers tree reconstruction, LPN combination, and final COT generation until a later processing interval. 
This transformation exposes receiver extension as a decoupled, stream-aligned stage without modifying the interactive protocol.

\textbf{Role-asymmetric expansion dependencies.}
Expanding the full set creates a long-lived intermediate representation: expanded tree outputs are generated during preprocessing but remain unused until their corresponding correlations are consumed. Deferring this local computation shortens the lifetime of the expanded representation and allows the compact communication-dependent information to be retained instead. 
However, the feasibility of deferral is not the same for sender and receiver OT roles. Each COT extension epoch processes $T$ GGM trees of height $h$ and combines their outputs with an LPN operand derived from the preceding extension state. The sender must expand every tree during preprocessing because the level-wise protocol messages and per-tree correlation term depend on the expanded values. Postponing sender expansion would delay the interactive exchange, while discarding and regenerating the expanded values would duplicate computation. In contrast, once the interactive exchange completes, the receiver has all information required for local tree reconstruction and requires no further input from the peer. We therefore retain the sender's required expansion but defer receiver-side reconstruction and subsequent local processing until closer to COT consumption.

\textbf{Compact representation.}
As in the conventional extension, the receiver performs the same GGM tree reconstruction and LPN computation;
our deferral enables retaining the expanded intermediate state in a compact form. 
For a tree of height $h$ and correlation width $\lambda$, the receiver's compact record contains $h$ selected $\lambda$-bit node values, $h$ selection bits, and one $\lambda$-bit correlation term. Therefore, the retained payload per tree for the conventional immediate (I) and our proposed deferred (D) expansion is 
$I(h,\lambda) = 2^h\lambda$ and $D(h,\lambda)= h\lambda + h + \lambda$
bits, respectively. The payload-level reduction is consequently
\begin{displaymath}
    \frac{I(h,\lambda)}{D(h,\lambda)}
    =
    \frac{2^h\lambda}
         {h\lambda + h + \lambda}.
    \label{eq:deferred_state_reduction}
\end{displaymath}
For $h=8$ and $\lambda=128$, the compact record occupies $152$\,B, compared with $4{,}096$\,B for the expanded leaf vector, reducing retained tree data by $26.9\times$.

\subsection{Cross-Phase Batch-Streaming Architecture}\label{Sec:design_batch}
The index-aligned COT batches exposed by Role-Aware Deferred COT Extension enable preprocessing and online execution to be reorganized as a cross-phase streaming dataflow. Rather than generating and retaining a complete correlation pool, the architecture produces, transfers, and consumes correlations at batch granularity. While the PS executes the input-dependent protocol for the current batch, the PL prepares the input-independent correlations required by a future batch. This organization introduces concurrency across the conventional offline--online boundary while ensuring that each batch is complete before its associated online operation begins.

\textbf{Architecture-aligned batching.} 
The batch size is derived jointly from the GGM-tree structure and the fixed geometry of the accelerator datapath, rather than from an arbitrary transfer size. A GGM tree of height $h$ yields $L_t=2^{h}$ leaf positions, while the engine delivers correlations in groups of $L_g = \beta\,(c_{\mathrm{btg}} + c_{\mathrm{on}})$ positions, where $\beta$ is the configured number of concurrent online consumers and $c_{\mathrm{btg}}, c_{\mathrm{on}}$ are the per-group triple and online consumption demand.
Since neither $L_t$ nor $L_g$ divides the other, a batch boundary can fall inside a tree. This enforces re-deriving any straddled tree in each batch that touches it and incurs repeated GGM work ($2.34\times$ the ideal number of expansions).
We therefore select the smallest batch containing an integral number of both trees and delivery groups: $L_b=\operatorname{LCM}(L_t, L_g)$. 

For the evaluated configuration ($n{=}2^{18}$, $h{=}8$, $\beta=n_t{=}4$, $c_{\mathrm{btg}}{=}11$, $c_{\mathrm{on}}{=}35$), $L_t=256$ and $L_g=184$, yielding a batch of $L_b=5{,}888$ positions, equivalent to $23$ trees and $32$ delivery groups. 
The selected batch then serves as the common unit of expansion, intermediate retention, and online consumption, while the cross-party message size remains independently configurable.

\textbf{Cross-phase pipeline.} 
Because correlation preparation is input-independent, the PL can process a future batch while the PS executes the input-dependent protocol for the current batch. 
After the initial pipeline fill, online execution of batch $j$ overlaps preprocessing of batch $j+\ell$, where $\ell$ is the bounded lookahead distance. 
The PS and PL thereby operate as concurrent consumer and producer stages, reducing the steady-state interval from the sum of their latencies toward that of the slower stage. Preprocessing latency is fully hidden only when the future batch completes before its consumption point; otherwise, the uncovered portion remains on the online critical path.

The architecture accommodates both full-epoch background preprocessing and fine-grained deferred processing through a common batch interface, while preserving each batch's logical indices, OT direction, and consumer assignment.

\subsection{Dual-path Correlation Processing}\label{sec:design_role}

Beaver triple generation consumes index-matched correlations from two opposite OT directions, requiring each party to process both sender-view and receiver-view data streams. 
Because time-multiplexing a single engine between roles introduces serialization stalls and limits throughput, we design a \emph{dual-path correlation-processing architecture}.
By provisioning dedicated, role-specific processing paths under a common delivery control, both OT directions required for Boolean triple generation advance concurrently without coarse-grained role reconfiguration.
This spatial separation enables the sender path to construct protocol messages concurrently with deferred expansion on the receiver path (Section~\ref{Sec:design_deferral}), while isolating their correlation state.

\textbf{Correlation alignment and triple processing.}
Beaver triple processing aligns sender-view and receiver-view correlations from opposite OT directions by logical generation and index. Because the two streams have a fixed generation offset, a bounded alignment buffer retains the earlier stream until its matching counterpart becomes available. The aligned correlations are then combined into local Boolean-triple shares within the PL.

The generated COTs are split among online OT use, Beaver triple processing, and correlations required as the LPN seed table for the next extension epoch according to public protocol parameters. Correlations assigned to triple processing remain within the PL, while only completed triple shares and correlations required by the online protocol are transferred to the PS. This placement removes the need for keeping intermediate correlations used exclusively for triple formation on the PS while preserving their logical ordering and cross-party association.

\subsection{Dual-CRH Split-stream Reformulation}\label{sec:split}
Motivated by the computational asymmetry identified in Section~\ref{sec:motivation_role_asymmetry}, we reformulate the sender computation as two index-aligned CRH streams. Rather than merely replicating the CRH operator, the proposed transformation reorganizes the input layout and its processing into two parallel operators, thereby preserving branch-level parallelism across the sender computation.
  
For each OT index $i$, the sender computes
$\mathrm{pad}_0[i] = H_{\kappa(s,i)}(m_i),$ and 
$\mathrm{pad}_1[i] = H_{\kappa(s,i)}(m_i \oplus \Delta)$.
where $s$ is the session seed and $\kappa(s,i)$ derives the CRH key for index $i$. 
Existing implementations commonly represent the corresponding inputs in a single interleaved layout,
$$\text{Layout}_{\text{interleaved}} = [m_0,\; m_0 \oplus \Delta,\; m_1,\; m_1 \oplus \Delta,\; \dots].$$
Directly mapping this representation to parallel CRH units forces their outputs to converge at a shared FIFO or memory port and incur multiple-writer contention. Simultaneous results must then be arbitrated, packed into a wider word, or buffered before serialization.
The shared write boundary can therefore limit the aggregate processing rate even when both CRH units individually sustain an initiation interval of $II=1$.

We instead transform the interleaved computation into two persistent streams:
$$
\text{Layout}_{\text{split}} =
\begin{cases}
[m_0,\; m_1,\; m_2,\; \dots], \\
[m_0 \oplus \Delta,\; m_1 \oplus \Delta,\; m_2 \oplus \Delta,\; \dots].
\end{cases}
$$
Each stream is assigned a dedicated CRH unit and an independent output FIFO, eliminating branch arbitration at the CRH write boundary.
Both CRH units are initialized with the same session seed, and advance their local index counters synchronously from the same origin. 
Consequently, the two streams preserve the index correspondence between
the correlated inputs and produce the same indexed pads as the reference
computation.


\subsection{Protocol Preservation}\label{sec:semantic_preservation}
The transformations modify only scheduling and dataflow. Deferred extension changes when a receiver-side COT becomes available, while preserving its value, index, order, and preprocessing messages. Split-stream processing changes how the sender-side branch hashes are performed after availability, while preserving the mapping among each OT index, correlated branch, CRH key, and output pad. Their shared interface therefore presents the same indexed outputs to the downstream protocol as the reference execution. Reformulations depend only on public protocol parameters and the known inference sequence, introducing no secret-dependent control flow.

\providecommand{\TODOwan}{\textcolor{red}{\texttt{[WAN]}}}


\section{Experiments and Results}\label{Sec:exp}

\subsection{Experimental Setup}
\subsubsection{\textbf{Platform Setting}} 
We implement our design using Vitis High-Level Synthesis~\cite{vitis_hls_user_guide}, targeting 170\,MHz and the Zynq UltraScale+ ZCU102.
The server runs on an Intel(R) Core(TM) i9-14900K Processor operating at 3.2 GHz.
We evaluate under two network settings: LAN (1\,Gbps, 0.3\,ms RTT) and WAN (320\,Mbps, 4\,ms RTT).

\begin{table}[t]
    \centering
    \caption{COT parameter configurations comparison for a workload of $5\times10^{8}$ COTs, each at a security level verified by~\cite{lpnsimulator2024}.}
    \label{tab:parameter_comparison}
    \vspace{-5pt}
    \resizebox{0.48\textwidth}{!}{%
    \begin{tabular}{@{} c c c c c c c c c c @{}}
    \toprule
    \multicolumn{5}{c}{\textbf{LPN/GGM Parameters}} & \textbf{On-Chip} & \multicolumn{2}{c}{\textbf{Execution}} & \multicolumn{2}{c}{\textbf{Per COT Cost}} \\
    \cmidrule(lr){1-5} \cmidrule(lr){7-8} \cmidrule(lr){9-10}
    \textbf{$n$} & \textbf{$h$} & \textbf{$t$} & \textbf{$k$} & \textbf{bits} & \textbf{(MB)} & \textbf{use/$n$} & \textbf{iters} & \textbf{AES} & \textbf{bits} \\
    \midrule
    \multirow{3}{*}{$2^{16}$}
    & 6  & 1,024 & 6,553   & 151.5 & 0.21 & 0.804 & 8,158  & 2.45 & 29.84 \\
    & 8  & 256   & 19,660  & 152.0 & 0.35 & 0.667 & 9,840  & 2.99 & 12.00 \\
    & 10 & 64    & 49,152  & 149.0 & 0.82 & 0.238 & 27,536 & 8.39 & 10.49 \\
    \midrule
    $2^{18}$ & 8  & 1,024 & 26,214  & 174.9 & 0.56 & 0.868 & 1,890 & 2.30 & 9.22  \\
    $2^{20}$ & 10 & 1,024 & 104,857 & 186.7 & 1.86 & 0.890 & 461   & 2.25 & 2.81  \\
    \midrule
    $10^{7}$ & 13 & 1,319 & 589,760 & 151.8 & 9.84 & 0.944 & 43    & 2.16 & 0.44  \\
    \bottomrule
    \end{tabular}%
    }
    \vspace{-10pt}
\end{table}
\begin{table}[t]
    \centering
    \footnotesize
    \caption{Microbenchmark comparison with $N = 200,704$ elements over LAN. 
    Percentages show contribution to total latency.}
    \label{tab:micro_comp}
    \vspace{-7pt}
    \resizebox{0.49\textwidth}{!}{%
    \setlength{\tabcolsep}{2.2pt} 
    \begin{tabular}{@{} p{1cm} c c c c c c @{}}
    \toprule
    \multirow{2}{*}{\textbf{}} & \multicolumn{2}{c}{\textbf{Protocol Execution}} & \multicolumn{2}{c}{\textbf{BTG}} & \multicolumn{2}{c}{\textbf{Total}} \\
    \cmidrule(lr){2-3} \cmidrule(lr){4-5} \cmidrule(lr){6-7}
     & \textbf{Base} & \textbf{Ours} & \textbf{Base} & \textbf{Ours} & \textbf{Base} & \textbf{Ours} \\
    \midrule
    SecComp   & 5.75 (66\%)  & \textbf{3.69} (95\%)  & 2.93 (34\%)    & \textbf{0.18} (4.7\%) & 8.68  & \textbf{3.88} (2.24$\times$) \\
    SecMult   & 0.04 (0.4\%) & \textbf{0.05} (0.6\%) & 11.36 (99\%)  & \textbf{8.15} (99\%)  & 11.41 & \textbf{8.20} (1.39$\times$) \\
    \bottomrule
    \end{tabular}
    }
    \vspace{-10pt}
\end{table}



\subsubsection{\textbf{Metrics and Evaluation Methodology}}
We evaluate the architecture across four core metrics: execution latency (seconds), nonlinear operator throughput ($\text{ops/s}$), memory footprint via PS peak Resident Set Size (MB), and PL hardware utilization using post-place-and-route reports with $36$\,Kb BRAM tile counts.

\subsubsection{\textbf{Baselines and Comparison Scope}}
To our knowledge, no public FPGA implementation targets the same OT-based 2PC nonlinear stack, making a direct comparison difficult without conflating architectural and protocol differences.
We therefore use protocol-matched baselines from the same protocol family.
The FPGA baseline (Table~\ref{tab:pl_breakdown}) preserves the baseline's bulk execution order while applying standard HLS optimizations and the COT generation method from~\cite{li2025silentflow}.
For transformer evaluation, we replace BumbleBee's RLWE-based \texttt{MulOLE} paths in GELU and Softmax with functionally equivalent OT-derived arithmetic-triple implementations and retain all other components.

\subsection{Design Configuration and Resource Utilization}\label{sec:config}
We explore the LPN parameter space to reduce off-chip seed accesses and fit the seed table within the available on-chip memory. The baseline 9\,MB table exceeds the client's shared L2 cache, increasing the LPN output-combination cost.
Unlike the baseline software configuration~\cite{huang2022cheetah}, which allocates $10^7$ correlation pools totaling $1{,}440$\,MB across eight generator instances, our selected parameters ($n{=}2^{18}$, $k{=}26{,}214$, $h{=}8$) allow the complete LPN seed table to be double-buffered in BRAM, using approximately $90\%$ of the available capacity.


Table~\ref{tab:pl_breakdown} summarizes the microarchitectural trade-offs between the two architectures. 
On-chip double buffering of the LPN seed table, together with deferred-expansion windows and per-lane stream buffers, avoids PS-side staging of intermediate data during each processing pass. The proposed design requires only $55$ DSPs, mainly for the arithmetic Beaver triple generation.

\textbf{Protocol Invariant Validation Methodology.}
We validate protocol invariants at four complementary levels.
First, under fixed randomness, deferred receiver expansion reproduces immediate extension outputs across all batch sizes. Second, multi-epoch execution verifies key state propagation across deferred epoch boundaries. 
Third, an immediate sender seamlessly pairs with a deferred receiver without changing correlation counts or logical consumption order. 
Fourth, sender--receiver outputs satisfy the required correlation relation at every vector index, while the split CRH streams reproduce the baseline-indexed pads.
Collectively, we establish bit-exact output equivalence and verify logical indexing, batch alignment, branch synchronization, and inter-epoch key propagation.


\subsection{Microbenchmarks}\label{sec:microbench}

\subsubsection{\textbf{Secure Comparison}}\label{sec:eval_comp}
As detailed in Table~\ref{tab:micro_comp}, preprocessing dominates the baseline execution: BTG accounts for $34\%$ of total latency, and COT generation alongside digit comparison consumes the remaining $66\%$. By offloading COT generation and BTG and overlapping with online execution, \textit{Interflow} achieves a $2.24\times$ speedup and reduces PS-side preprocessing overhead to a lightweight DDR transfer, which contributes only $5\%$ of total execution time. 
The remaining latency is therefore dominated by online protocol execution, showing that \textit{Interflow} successfully hides correlation generation from the critical path.

\begin{table}[t]
    \centering
    \caption{Comparison of hardware resource utilization breakdown.}
    \label{tab:pl_breakdown}
    \vspace{-5pt}
    \resizebox{0.48\textwidth}{!}{%
    \begin{tabular}{@{} l c c c c c c @{}}
    \toprule
    \multirow{2}{*}{\textbf{Component}} & \multicolumn{3}{c}{\textbf{Baseline Design}} & \multicolumn{3}{c}{\textbf{Proposed Design}} \\
    \cmidrule(lr){2-4} \cmidrule(lr){5-7}
     & \textbf{BRAM} & \textbf{FF} & \textbf{LUT} & \textbf{BRAM} & \textbf{FF} & \textbf{LUT} \\
    \midrule
    COT generator (S)   & 94.5 & 52,962 & 29,997 & 289.5 & 14,831 & 24,546 \\
    COT generator (R)   & 98.5 & 121,625 & 88,298 & 289.5 & 14,815 & 25,120 \\
    Boolean BTG         & \multicolumn{3}{c}{\textit{on PS}} & 83.5 & 20,396 & 17,813 \\
    Arithmetic BTG      & \multicolumn{3}{c}{\textit{on PS}} & 73.5 & 32,670 & 38,092 \\
    COT splitter        & \multicolumn{3}{c}{\textit{n/a}}   &  0.0 &    629 &    326 \\
    AXI control         & 47.5 &  35,055 &  19,986 &  81.5 & 50,274 & 31,142 \\
    \midrule
    Total               & 240.5 & 209,642 & 138,281 & 817.5 & 133,615 & 137,039 \\
    \bottomrule
    \end{tabular}
    }
    \vspace{-8pt}
\end{table}
\begin{table}[t]
    \centering
    \caption{Per-operator throughput improvement across network settings with $N=200,704$ elements.}
    \label{tab:nonlin_comp}
    \vspace{-5pt}
    \setlength{\tabcolsep}{4pt}
    \resizebox{0.49\textwidth}{!}{%
    \begin{tabular}{@{} l *{8}{c} @{}}
    \toprule
    \multirow{2}{*}{\textbf{Method}} & \multicolumn{2}{c}{\textbf{ReLU}} & \multicolumn{2}{c}{\textbf{MaxPool}} & \multicolumn{2}{c}{\textbf{GELU}} & \multicolumn{2}{c}{\textbf{Softmax}} \\
    \cmidrule(lr){2-3} \cmidrule(lr){4-5} \cmidrule(lr){6-7} \cmidrule(lr){8-9}
     & \textbf{LAN} & \textbf{WAN} & \textbf{LAN} & \textbf{WAN} & \textbf{LAN} & \textbf{WAN} & \textbf{LAN} & \textbf{WAN} \\
    \midrule
    Base & 33{,}540 & 32{,}371 & 3{,}028 & 3{,}001 & 5{,}276 & 5{,}076 & 2{,}988 & 3{,}170 \\
    Ours & \textbf{51{,}946} & \textbf{48{,}961} & \textbf{4{,}494} & \textbf{4{,}313} & \textbf{7{,}382} & \textbf{7{,}046} & \textbf{4{,}428} & \textbf{4{,}099} \\
    Gain & 1.55$\times$ & 1.51$\times$ & 1.48$\times$ & 1.44$\times$ & 1.40$\times$ & 1.39$\times$ & 1.48$\times$ & 1.29$\times$ \\
    \bottomrule
    \end{tabular}
    }
    \vspace{-6pt}
\end{table}

\subsubsection{\textbf{Secure Multiplication}}\label{sec:eval_mult}
As shown in Table~\ref{tab:micro_comp}, 
arithmetic Beaver triple generation accounts for $99.6\%$ of the baseline secure-multiplication latency. 
The offline generator uses Gilboa's OT-based multiplication protocol~\cite{gilboa1999} to construct the two cross terms required by each arithmetic triple,
consuming $2\ell$ correlations per general multiplication triple, or $64$ COTs at $\ell=32$.
\textit{Interflow} achieves a $1.39\times$ speedup at $\ell=32$, rising to $1.45\times$ at $\ell=64$ as the correlation demand doubles. 

\subsection{Nonlinear Operation Evaluation}

\begin{figure}[t] 
    \centering 
    \includegraphics[width=\linewidth, trim={0.1cm 0.1cm 0cm 0cm}, clip ]{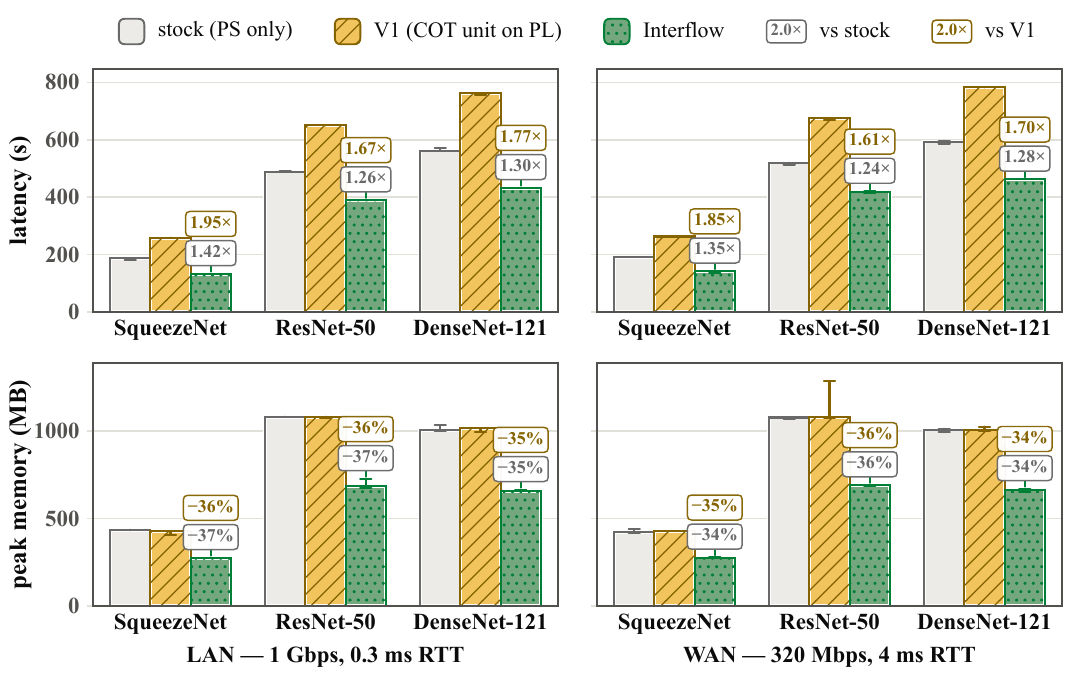} 
    \vspace{-20pt}
    \caption{Latency and memory comparison of end-to-end secure inference for the software baseline, the protocol-matched FPGA baseline, and \textit{Interflow}.} 
    \label{fig:e2e_comp} 
    \vspace{-13pt}
\end{figure}

For the non-linear operators in Table~\ref{tab:nonlin_comp}, \textit{Interflow} achieves throughput speedups of $1.51{-}1.55\times$ for ReLU, $1.44{-}1.48\times$ for MaxPooling, $1.4\times$ for GELU, and $1.29{-}1.48\times$ for Softmax, compared to the protocol-matched software baseline across the LAN and WAN settings. 
The gains vary with each operator's correlation demand and the fraction of execution exposed to preprocessing.
\textit{Interflow} removes the 
critical-path preprocessing stall present in the baseline and bounds the resident correlation footprint at $28.6$\,MB for all operators. In contrast, the baseline retains correlation pools whose capacity grows with the extension-epoch size, from $40.1$\,MB at $n{=}2^{18}$ to $1{,}440$\,MB at $n{=}10^{7}$.

\subsection{End-to-end Inference}
Fig.~\ref{fig:e2e_comp} summarizes end-to-end secure inference results. Under the LAN network setting, \textit{Interflow} achieves speedups of $1.42\times$, $1.26\times$, and $1.30\times$ over the software baseline, and $1.95\times$, $1.67\times$, and $1.77\times$ over the protocol-matched FPGA baseline across SqueezeNet, ResNet-50, and DenseNet-121, respectively.
It also reduces peak client memory by up to $37\%$ and server memory by $200$--$300$\,MB, with consistent reductions across network settings due to bounded correlation buffering.
In contrast, the execution speedup decreases under WAN latency.
The results also show that the protocol-matched baseline executes slower than software ($0.73$--$0.75\times$) at the comparable memory footprint. 
Accelerating correlation generation alone does not prevent refill stalls when complete epochs are returned to memory.
\emph{Interflow} instead delivers batch-level correlations, keeping correlations available to the online phase.

Under LAN and WAN settings, \textit{Interflow} accelerates the nonlinear block by $1.44\times$ and $1.34\times$, respectively, yielding end-to-end speedups of $1.21\times$ and $1.17\times$ for a single-layer BERT-base model. The smaller end-to-end gains reflect the small contribution of nonlinear blocks to the baseline inference time.

\section{Conclusion}\label{Sec:conclusion}
This paper presented \emph{Interflow}, an algorithm--hardware co-design that coordinates correlation preprocessing and online execution for OT-based secure nonlinear inference. 
Interflow resolves system bottlenecks by decoupling preprocessing at its final communication dependency and executing in a unified, role-aware datapath. 
Evaluated on the Zynq UltraScale+ ZCU102, Interflow improves nonlinear operator throughput by up to $1.55\times$ for CNN operators and $1.48\times$ for transformer operators, yielding geometric-mean end-to-end speedups of $1.28\times$ over the protocol-matched software baseline and $1.72\times$ over an FPGA baseline without triple generation offload, while reducing the client memory footprint by up to $37\%$.
These results demonstrate that managing the complete correlation lifecycle is essential for efficient secure edge inference.

{\footnotesize 
\bibliographystyle{IEEEtran}
\bibliography{cite.bib}
}

\end{document}